# Piezoacoustic wave spectra using improved surface impedance matrix: Application to high impedance-contrast layered plates

Victor Y Zhang, Bertrand Dubus,
Bernard Collet and Michel Destrade

2007

Starting from the general modal solutions for a homogeneous layer of arbitrary material and crystalline symmetry, a matrix formalism is developed to establish the semi analytical expressions of the surface impedance matrices (SIM) for a single piezoelectric layer. By applying the electrical boundary conditions, the layer impedance matrix is reduced to a unified elastic form whether the material is piezoelectric or not. The characteristic equation for the dispersion curves is derived in both forms of a 3-dimentional acoustic SIM and of an electrical scalar function. The same approach is extended to multilayered structures such as a piezoelectric layer sandwiched in between two metallic electrodes, a Bragg coupler, and a semi-infinite substrate as well. The effectiveness of the approach is numerically demonstrated by its ability to determine the full spectra of guided modes, even at extremely high frequencies, in layered plates comprising of up to four layers and three materials. Negative slope in *f-k* curve for some modes, asymptotic behavior at short wavelength regime, as well as wave confinement phenomena made evident by the numerical results are analyzed and interpreted in terms of the surface acoustic waves and of the interfacial waves in connection with the bulk waves in massive materials.

## I. INTRODUCTION

Engineering applications in surface acoustic wave (SAW) devices, composite material characterization, and smart structures, require the analysis of acoustic wave interaction with anisotropic and/or piezoelectric multi-layers. Recently, the analysis of *bulk acoustic waves* (BAW) in multi-layered structures composed of stacked piezoelectric, dielectric, and metallic materials has once again attracted the attention of engineers and researchers working on radio frequency components such as stacked crystals filter, coupled resonator filter, and multiplexers built on *solidly mounted resonators* (SMR) and thin film bulk acoustic resonators. The development of efficient simulation tools is needed to characterize accurately the electromechanical behaviour of complex stratified structures accounting for realistic electrical and mechanical interface and *boundary conditions* (BC).

Many methods have been developed for studying the acoustic waves spectra in layered media, and a majority of them are based on the matrix formalism. The most commonly known is without any doubt the *transfer matrix method* (TMM),[1] which in its simplest form is conceptually intuitive but intrinsically suffers from numerical instability at high frequencies. Following many efforts to improve the TMM,[2-4] a genuine advance in breaking through the numerical limitation of the classical TMM was the approach using the *surface impedance matrix* (SIM).[5-18] Although the concept of the acoustical (or mechanical) impedance is an old one for all physicists and engineers, the SIM expressed in its initial form[13] did not guarantee numerical stability at high frequencies because the exponential dichotomy factors generating instability were not arranged in a convenient way. By means of a reformulated form of the SIM[9,11,12,14,16,17] and its equivalent variants - reflection matrix,[19] scattering matrix,[20,21] or compliance/stiffness matrix,[22,23] for individual layers and with the help of a recursive algorithm for the overall multilayers,[9,11,12,16,17,19-23] correct transitions of the state vector across layer interfaces can be achieved. This in turn allows numerically stable and robust algorithms to be elaborated without increasing the basic matrix size.

The SIM approach consists in first, expressing the state vector values at the two surfaces of a piezoelectric layer by means of an 8-dimensional layer impedance matrix; then, calculating the global surface impedance matrix of a multilayer with the help of a recursive algorithm from the impedance matrix of the individual layers along with the imposed/known BC; finally, formulating the characteristic equations of guided waves by cancelling the determinant of the final impedance matrix formulated for a selected interface where the acoustic energy is concentrated. The key point in calculating the layer impedance matrix, which guarantees the numerical stability, is the appropriate arrangement of the eight partial modes resulting from the constitutive relations and the equilibrium equations. Common to various possible forms, it is essential to avoid exponential dichotomy by absorbing the exponentially large terms through the negligibly small amplitudes,[17] both of which are associated with the same partial modes. The advantages of the SIM approach over the well-known TMM approach are analyzed in some detail in many published works. The approach, based on the decomposition of incident and reflected partial waves at the different interfaces, acts on the amplitudes of partial modes of each layer rather than on the physical quantities themselves. Thus as a result of energy conservation, such an approach is guaranteed to be stable.[19] The loss of precision in the SIM is delineated in terms of upward-bounded and downward-bounded waves. The resultant expressions and algorithm are terse so that their implementation is more convenient and efficient than the TMM. This alternative formulation is stable, efficient, and illuminating,[21] while being as fast as the TMM.[16] The matrix form is concise, thus simple to program and implement.[22] The two concise recursive algorithms, based on combining Stroh formalism and a SIM approach, are more stable than the standard matrix method and are faster than the global matrix method. They also appear to be extremely straightforward and efficient at least from the computational point of view.[17] The SIM approach also has the advantage of being conceptually simple and numerically flexible. Namely, a SIM can be defined for any interface as well as for an external surface when the structure is a multilayer. This flexibility allows one to numerically determine the dispersion curves and field distributions with more accuracy by selecting the appropriate interface in the neighbourhood of which the electromechanical energy flux is localized, and to obtain the full set of solutions, even under extreme conditions, by repeatedly formulating the SIM at different locations. The modeling of Lowe[1,24] for ultrasonic waves in multilayered media is also restricted to isotropic and purely elastic materials. Stewart and Yong[25] performed an analysis of the acoustic propagation in multilayered piezoelectric plates using the TMM. Both simple thickness modes and general dispersion behaviors for propagating straight-crested waves in zinc oxide on silicon thin film resonators were studied. However, only the mass and no stiffness effects of the electrodes on the resonators were considered in their model. Acoustical spectra become very complex in multi-layered plates composed of layers having significantly different material properties, especially when the constituent layers have high velocity and/or impedance contrast. With a slight variation of the pair $\omega$-k value in the spectral domain, where $\omega$ is the frequency and k is the wave number, the nature of the guided modes can be radically different, from Lamb-like to SAW-like.

In this paper, we develop a detailed formalism enabling the full spectra of guided waves in layered plates to be calculated even with a high impedance-contrast and at extremely high frequencies. The exemplified structure consists of a piezoelectric AlN layer, a high-impedance metallic tungsten (W) layer, and a low-impedance dielectric $SiO_2$. This study was motivated by the SMR design employing AlN as active resonator and the stacked cells made of bi-layer $W/SiO_2$ as Bragg coupler, to isolate the resonator from the substrate. We develop a semi-analytical model to calculate the dispersion curves for various multi-layer configurations. Starting from an adaptation of the general Stroh formalism[26,27] to the special case of homogeneous materials for every layer,[7,10,11,28,29] we recall briefly in Sec. II the main results and formulas necessary for further developments. SIM expressions and dispersion relations in terms of the SIM elements are derived in Sec. III first for a single piezoelectric layer and then extended to include two surrounding metallic electrodes and a semi-infinite supporting substrate via a Bragg coupler. Section IV is devoted to numerical investigations considering only a piezoelectric AlN layer combined with a few $W/SiO_2$ cells in order to facilitate the analysis of the rather complicated spectra of guided acoustic modes. Some conclusions and discussions are given in Sec. V.

## II. BASIC STROH FORMALISM AND MAIN RESULTS

To simplify the formulation, we choose the reference coordinate system such that the $x_1$–axis is the propagation direction in the layering plane, the $x_2$–axis is parallel to the layering thickness, and the layering plane is assumed to be unbounded. The vibration state of a structure can be described by means of a state vector ($\boldsymbol{\tau}$) whose components are composed of physical variables that are continuous across the interfaces.[27-29] The state vector we chose is defined by $\boldsymbol{\tau}=[\ T_{21}\ T_{22}\ T_{23}\ D_2\ v_1\ v_2\ v_3\ \psi\ ]^T$, where $v_i$ are the components of particle velocity vector $\boldsymbol{v}$, $T_{2i}$ are the components of stress tensor in $x_2$-plane; $\psi \equiv j\omega\phi$; $j^2 \equiv -1$, $\phi$ is the electric potential, $\omega$ is the angular frequency, and $D_2$ is the normal electric displacement. The superscript $T$ means transpose, and the subscript i (=1, 2, 3), corresponds to the space variable $x_i$. This choice of the state vector is identical to the choice in Ref. 28, but different from that adopted in Refs. 9-11 where the displacement was employed instead of the velocity and where the components of $\boldsymbol{\tau}$ were not arranged in the same order. According to results established in Refs. 28, 29 and 13, the state vector $\boldsymbol{\tau}$ obeys the following system of ordinary differential equation of first order in the harmonic regime,

$$\frac{d\boldsymbol{\tau}}{dx_2} = j\omega\mathbf{A}\boldsymbol{\tau}\ . \tag{1}$$

The system matrix $\mathbf{A}$, also named the state matrix or Stroh matrix, is given by

$$\mathbf{A} = \begin{bmatrix} \boldsymbol{\Gamma}_{12}\boldsymbol{\Gamma}_{22}^{-1}s_1 & \left(\boldsymbol{\Gamma}_{12}\boldsymbol{\Gamma}_{22}^{-1}\boldsymbol{\Gamma}_{21} - \boldsymbol{\Gamma}_{11}\right)s_1^2 + \boldsymbol{\rho}_0 \\ \boldsymbol{\Gamma}_{22}^{-1} & \boldsymbol{\Gamma}_{22}^{-1}\boldsymbol{\Gamma}_{21}s_1 \end{bmatrix}, \text{ with } \boldsymbol{\Gamma}_{ab} = \begin{bmatrix} c_{1a1b} & c_{1a2b} & c_{1a3b} & e_{b1a} \\ c_{2a1b} & c_{2a2b} & c_{2a3b} & e_{ba} \\ c_{3a1b} & c_{3a2b} & c_{3a3b} & e_{b3a} \\ e_{a1b} & e_{a2b} & e_{a3b} & -\varepsilon_{ab} \end{bmatrix}$$

(a,b=1, 2, 3),

where $\boldsymbol{\rho}_0$ is $\rho$ times a 4-dimensional identity matrix with a $4^{th}$ element equal to zero, and c, e, $\varepsilon$, and $\rho$ denote material constants. It is worthwhile recalling that the material constants appearing in the matrices $\boldsymbol{\Gamma}_{ab}$ are those evaluated in the working reference system. The matrix $\mathbf{A}$ has a frequency-independent form. It depends, apart from the material constants, on a unique variable $s_1$, with $s_i = k_i/\omega$, $k_i$ being the component along $x_i$ of the wave number, and $s_i$ the slowness component. It is from the $\mathbf{A}$ matrix that result the proper modes, also called partial waves, susceptible of propagating in a homogeneous unbounded medium. According to the linear system theory, the general solution of $\boldsymbol{\tau}$ can be constructed by a linear combination of modal solutions weighted with modal amplitude ($\boldsymbol{y}$). This yields the general form for the state vector

$$\boldsymbol{\tau}\,(x_1, x_2, t) = \mathbf{Q}\mathbf{E}(x_2)\boldsymbol{y}\,e^{j\omega(t - s_1 x_1)}\ , \tag{2}$$

where

$$\mathbf{E}(x_2) = e^{-j\omega\boldsymbol{s_2}x_2} \tag{3}$$

is the transition matrix, also called the "propagator matrix", and $\boldsymbol{s_2}$ is the diagonal spectral matrix of (1), $\mathbf{Q}$ is the associated modal matrix, and $\boldsymbol{y}$ is the amplitude vector. The components $s_2^{(r)}$ of $\boldsymbol{s_2}$ and $\mathbf{Q}^{(r)}$ of $\mathbf{Q}$ (r = 1,...,8), are eigenvalues and eigenvectors determined from the following homogeneous system

$$[\mathbf{A} + s_2^{(r)}\mathbf{I}]\mathbf{Q}^{(r)} = \mathbf{0},\ r = 1,...,8, \tag{4}$$

where $\mathbf{I}$ is a 8-dimensional identity matrix. In what follows, we assume that the partial modes are arranged in such a manner that the matrices $\boldsymbol{s_2}$ and $\mathbf{Q}$ can be put into the form:

$$\mathbf{s}_2 = \begin{bmatrix} \mathbf{s}_D & \mathbf{0} \\ \mathbf{0} & \mathbf{s}_I \end{bmatrix}, \; \mathbf{Q} = \begin{bmatrix} \mathbf{t}_D & \mathbf{t}_I \\ \mathbf{v}_D & \mathbf{v}_I \end{bmatrix}, \tag{5}$$

where all sub-matrices are 4-dimensional.

Hereafter, the subscripts $D$ and $I$ stand for *direct* and *inverse* partial waves, respectively, see Ref. 29 for their definition. Note that in Ref. 29 the words "diffracted" and "incident" were used in the place of "direct" and "inverse", respectively, in the context on surface wave problems. We point out that the real $\mathbf{s}_2$ s represent the upward- and downward-propagating oblique plane waves in the layers (propagating in $x_1$ and standing in $x_2$), and that their interaction is the origin of the resultant guided modes pertaining to the layered plate. We remind that all eight partial modes are required for a finite thickness layer, whereas only the 4 $D$- or $I$-modes are needed for a semi-infinite substrate according to the Sommerfeld radiation condition. In this paper, the case $s_1=0$, which corresponds to normal propagation, will not be considered, and the rare degenerated cases where the eigenvalue (or Christoffel equation's root) $s_2$ is double-valued for certain isolated special values of $s_1$ are also excluded, so that (2) is used as the valid solution throughout the paper.

## III.  MATRIX  FORMALISM  FOR  GUIDED  WAVES  IN  LAYERED STRUCTURES

We first consider a plate composed of three materials, namely a piezoelectric {c}-oriented AlN layer, a metallic W layer, and a purely dielectric $SiO_2$ layer, as illustrated by Fig. 1. In order to apply the results of the previous Section and to develop the matrix formalism for the purpose of deriving the characteristic equation leading to the dispersion relation of acoustic waves in this structure, we introduce some specific quantities for the values of the relevant physical quantities at the layer surfaces. In Fig. 1 the superscript "−" refers to the upper surface and the superscript "+" to the bottom surface , $\sigma^\pm$ stands for surface charge density, $V$ is the voltage across the piezoelectric layer, and $J$ is the current entering into the piezoelectric layer when an external load impedance $Z_L$ is connected between two surface electrodes. These latter are assumed for the moment to be perfectly conducting and negligibly thin so that they have no mechanical effects on the piezoelectric resonator. The effects of actual metallic electrodes of finite thickness are discussed in Sec. III-D.

FIG. 1.

### A. Single piezoelectric layer

We first look for expressions of the characteristic impedance matrices in terms of the modal ($\mathbf{Q}$) and spectral ($\mathbf{s}_2$) matrices for a piezoelectric layer. For this purpose, we introduce a *generalized mixed  matrix* $\mathbf{G}$ defined by

$$\begin{bmatrix} \mathbf{T}^- \\ \mathbf{T}^+ \end{bmatrix} \equiv [\mathbf{G}] \begin{bmatrix} \mathbf{V}^- \\ \mathbf{V}^+ \end{bmatrix}, \tag{6}$$

with $\mathbf{T}^\pm \equiv \begin{bmatrix} \mathbf{t}^\pm & D_2^\pm \end{bmatrix}^T$, $\mathbf{V}^\pm \equiv \begin{bmatrix} \mathbf{v}^\pm & \psi^\pm \end{bmatrix}^T$, $\mathbf{t} \equiv [t_{21} \; t_{22} \; t_{23}]^T$, and $\mathbf{v} \equiv [v_1 \; v_2 \; v_3]^T$. After (2), (3), and (5), we derive an expression of $\mathbf{G}$-matrix as follows

$$\mathbf{G} = \begin{bmatrix} \mathbf{t}_D & \mathbf{t}_I (\mathbf{E}_I^0)^{-1} \\ \mathbf{t}_D \mathbf{E}_D^0 & \mathbf{t}_I \end{bmatrix} \begin{bmatrix} \mathbf{v}_D & \mathbf{v}_I (\mathbf{E}_I^0)^{-1} \\ \mathbf{v}_D \mathbf{E}_D^0 & \mathbf{v}_I \end{bmatrix}^{-1}, \tag{7}$$

where $\mathbf{E}_{D,I}^0 = e^{-j\omega\mathbf{s}_{D,I}h}$ is the sub-matrix of the transition matrix $\mathbf{E}^0 \equiv \mathbf{E}(x_2 = h)$. We point out that other forms of **G**-matrix exist.[17,23] Each of the four sub-matrices $\mathbf{G}_{ij}$, $i$, $j = 1,2$, of **G** is 4-dimensional and contains elements of different physical signification. Take $\mathbf{G}_{11}$ for example and rewrite it in the form

$$\mathbf{G}_{11} = \begin{bmatrix} \mathbf{Z}_{11}^p & \mathbf{K}_{11} \\ \mathbf{X}_{11} & Y_{11} \end{bmatrix} \equiv \begin{bmatrix} \mathbf{Z}_m & \mathbf{K} \\ \mathbf{X} & Y_e \end{bmatrix}. \tag{8}$$

In (8), $\mathbf{Z}_m$ is a 3-by-3 matrix representing a mechanical impedance, $Y_e$ is a scalar proportional to an electrical admittance, while **K** and **X** have the dimension of a piezoelectric coefficient. $\mathbf{X} = \mathbf{K}^T$ provided an appropriate normalization is operated for the state vector. We emphasize that $\mathbf{Z}_m$ itself already includes a part originated from piezoelectricity. $\mathbf{G}_{11}$ is the counterpart to the *generalized surface impedance* matrix involved in the SAW problem.[29] When the layer material is non piezoelectric (dielectric or metallic, say), $\mathbf{K} = \mathbf{0}$ holds and $\mathbf{G}_{11}$ is decoupled into two independent parts, namely $\mathbf{Z}_m$ for the acoustic part and $Y_e$ for the electric part. This means that the electrical and mechanical variables can be dealt with independently. This happens equally to a piezo-inactive acoustical mode in piezoelectric materials. The same is true for the other sub-matrices of **G** which have analogous properties. Some interesting relations exist linking these sub-matrices. For a metallic layer, we only take into account the mechanical effects by assuming it to be perfectly conductive. We conclude by saying that (7) naturally applies for a non-piezoelectric layer.

In order to keep the matrix dimension-compatibility and to facilitate the transition at interfaces within a heterostructure, we find it more useful to define the *acoustic impedance* (**Z**) matrix as follows

$$\begin{bmatrix} \mathbf{t}^- \\ \mathbf{t}^+ \end{bmatrix} \equiv [\mathbf{Z}] \begin{bmatrix} \mathbf{v}^- \\ \mathbf{v}^+ \end{bmatrix}. \tag{9}$$

**Z** is 6-dimensional and appears in a purely elastic form. However, it implicitly includes the dielectric and piezoelectric effects when the layer is piezoelectric. This new definition applies whether the layer is piezoelectric or not.

## B. Electrical BC and characteristic equations

We now express the **Z** matrix defined in (9) in terms of the elements of the matrix **G** by eliminating the electrical variables upon applying the known electrical BC. The surface charge is written as $\sigma^- = D^-_2 - D^+_0$ and $\sigma^+ = D^-_0 - D^+_2$ at the upper and bottom surfaces, respectively. Relations between the electric potential and displacement $D^\pm_0$ in the vacuum side are established, after having resolved the Laplace equation, as $D^+_0 = g\psi^-$ and $D^-_0 = -g\psi^+$ with $g \equiv j|s_1|\varepsilon_0$. We deduce from them that

$$\sigma^- = D^-_2 - g\psi^- \tag{10a}$$

and that

$$\sigma^+ = -[D^+_2 + g\psi^+]. \tag{10b}$$

The charge conservation reads $\sigma^- + \sigma^+ = 0$, which gives

$$\sigma^- = -\sigma^+ \equiv \sigma. \tag{11}$$

Two types of electrical BC are to be distinguished for a piezoelectric layer :

1) the surface is metallized and short-circuited (SC), in which case $Z_L$=0, leading to

$$j\omega V \equiv \psi^- - \psi^+ = 0. \tag{12}$$

2) the surface is non-metallized and open-circuited (OC), in which case $Z_L$=∞, leading to

$$\sigma = 0, \text{ with } \psi^- \neq \psi^+ \text{ in general.} \tag{13}$$

In order to make the subsequent development easier, and especially in order to facilitate the introduction of the BC, we rewrite the relation (6) defining the **G**-matrix more explicitly and in a slightly different form:

$$
\begin{bmatrix} \mathbf{t}^- \\ \mathbf{t}^+ \\ \sigma \\ \sigma \end{bmatrix}
\equiv
\begin{bmatrix}
\mathbf{Z}_{11}^p & \mathbf{Z}_{12}^p & \mathbf{K}_{11} & \mathbf{K}_{12} \\
\mathbf{Z}_{21}^p & \mathbf{Z}_{22}^p & \mathbf{K}_{21} & \mathbf{K}_{22} \\
\mathbf{X}_{11} & \mathbf{X}_{12} & Y_{11}-g & Y_{12} \\
\mathbf{X}_{21} & \mathbf{X}_{22} & Y_{21} & Y_{22}+g
\end{bmatrix}
\begin{bmatrix} \mathbf{v}^- \\ \mathbf{v}^+ \\ \psi^- \\ \psi^+ \end{bmatrix}.
\tag{14}
$$

(10a,b) and (11) have been taken into account in arriving at (14). The upper-left sub-block in (14) is a mechanical impedance, the lower-right sub-block is an electrical admittance with the term *g* coming from the vacuum contribution. The anti-diagonal sub-blocks represent the piezoelectric terms.

Since the AlN layer in a resonator configuration is usually sandwiched between two metallic electrodes, we consider mainly the case where both surfaces are SC and we determine the characteristic equations corresponding to this condition. Electric SC condition implies $V$=0, i.e., $\psi^- = \psi^+$. Since the value of the potential is relative to a certain reference arbitrarily chosen, it is usual to set them both equal to zero. Use of $\psi^- = \psi^+ = 0$ in (14) yields, according to (9), the **Z**-matrix for a SC piezoelectric layer :

$$
\begin{bmatrix} \mathbf{t}^- \\ \mathbf{t}^+ \end{bmatrix}
= \begin{bmatrix} \mathbf{Z}^{sc} \end{bmatrix}
\begin{bmatrix} \mathbf{v}^- \\ \mathbf{v}^+ \end{bmatrix}, \text{ with } \mathbf{Z}^{sc} \equiv
\begin{bmatrix} \mathbf{Z}_{11}^p & \mathbf{Z}_{12}^p \\ \mathbf{Z}_{21}^p & \mathbf{Z}_{22}^p \end{bmatrix}.
\tag{15}
$$

$\mathbf{Z}^{sc}$ is the layer mechanical impedance matrix when its surfaces are both short-circuited (superscript SC). By further applying the stress-free mechanical BC, say $\mathbf{t}^- = \mathbf{0}$ at the upper surface, we finally arrive at the so-called *surface impedance matrix* (SIM) $\mathbf{Z}^+$ defined for the bottom surface as

$$\mathbf{t}^+ \equiv \mathbf{Z}^+ \mathbf{v}^+, \text{ with } \mathbf{Z}^+ = \mathbf{Z}_{22}^{sc} - \mathbf{Z}_{21}^{sc}(\mathbf{Z}_{11}^{sc})^{-1}\mathbf{Z}_{12}^{sc}. \tag{16a}$$

In the same way, we can define the SIM $\mathbf{Z}^-$ for the top surface by applying SC and stress-free BC at the bottom surface. The result is, after interchanging the matrix indices 1 and 2 in (16a),

$$\mathbf{t}^- \equiv \mathbf{Z}^- \mathbf{v}^-, \text{ with } \mathbf{Z}^- = \mathbf{Z}_{11}^{sc} - \mathbf{Z}_{12}^{sc}(\mathbf{Z}_{22}^{sc})^{-1}\mathbf{Z}_{21}^{sc}. \tag{16b}$$

$\mathbf{Z}_{ij}^{sc}$ above refer to the sub-matrices of $\mathbf{Z}^{sc}$. The SIM $\mathbf{Z}^\pm$ defined in (16) are the acoustic impedances looking outward from one surface toward and accounting for the mechanical BC imposed upon the other surface along with the electrical BC applied at both surfaces. The solutions of plate modes in general, and of the Lamb and SH modes in particular, in a piezoelectric layer with SC and stress-free at both surfaces are then obtained by simply canceling the $\mathbf{Z}^\pm$ determinant:

$$[\mathbf{Z}^\pm]\mathbf{v}^\pm = \mathbf{0} : \mathbf{v}^\pm \neq \mathbf{0} \text{ and } \Delta^{sc}(\omega, k_1) \equiv det[\mathbf{Z}^\pm] = 0 \implies k_{sc}(\omega). \tag{17}$$

Any one of the couple sign ± can be used and it results, in principle, in identical solutions. However, the functional behavior of $det[\mathbf{Z}^\pm(\omega, k_1)]$ is different in case of a multi-layer and especially at high frequencies. This has practical consequences in numerical resolution.

Once the values of $k_1$ are found for a given $\omega$ (usually by means of an iterative algorithm), the corresponding modes polarization can be determined from (17). The form of the characteristic equation in (17) has the advantage of easily separating the sagittal plane (SP) modes ($|\mathbf{Z}^{\pm}(1{:}2,1{:}2)|{=}0$) from the shear horizontal (SH) modes ($\mathbf{Z}^{\pm}(3,3){=}0$) when they are uncoupled. Such is the case for acoustic modes in the layered structure depictured by Fig. 1. The dispersion relation of plate modes is a plot of $\omega$ versus $k_1$ over a certain range with like modes following regular curves, which gives a graphical picture of the solutions of the wave equation for a (specific layered) plate. There is an infinite number of $\omega$ values which can satisfy (17) for a given $k_1$. But there will be in general a finite number of real $k_{sc}$ that can be found for a given $\omega$, and that number increases with $\omega$. Of course, the situation becomes complicated, if the solutions are not restricted to within the real domain. At a given real frequency, besides the propagating modes which are associated with the real solutions of $k_1$, there are a finite number of non-propagating modes having purely imaginary wave-numbers and an infinite number of inhomogeneous modes having complex wave-numbers.[32] In this paper, we limit our investigation to the propagating plate modes. Most of the solutions exist only for a frequency larger than a critical threshold (cut-off). The cut-off frequencies of the propagating plate modes in the limit of $k_1 {\to} 0$ represent nothing but the resonant frequencies of thickness modes which are the main waves for BAW devices.

The $\mathbf{Z}$-matrix associated to the OC condition can be determined by simply canceling out the electrical charge in (14). With $\sigma{=}0$, we obtain the counterpart to (15) as

$$\begin{bmatrix} \mathbf{t}^- \\ \mathbf{t}^+ \end{bmatrix} \equiv \begin{bmatrix} \mathbf{Z}^{oc} \end{bmatrix} \begin{bmatrix} \mathbf{v}^- \\ \mathbf{v}^+ \end{bmatrix}, \text{ with } \mathbf{Z}^{oc} = \mathbf{Z}^{sc} - \mathbf{K}\mathbf{Y}^{-1}\mathbf{X}. \tag{18a}$$

The matrices $\mathbf{K}$, $\mathbf{X}$, and $\mathbf{Y}$ involved in the $\mathbf{Z}^{oc}$-expression are defined by:

$$\mathbf{K} \equiv \begin{bmatrix} \mathbf{K}_{11} & \mathbf{K}_{12} \\ \mathbf{K}_{21} & \mathbf{K}_{22} \end{bmatrix}, \ \mathbf{X} \equiv \begin{bmatrix} \mathbf{X}_{11} & \mathbf{X}_{12} \\ \mathbf{X}_{21} & \mathbf{X}_{22} \end{bmatrix}, \text{ and } \mathbf{Y} \equiv \begin{bmatrix} Y_{11}-g & Y_{12} \\ Y_{21} & Y_{22}+g \end{bmatrix}. \tag{18b}$$

Use of $\mathbf{Z}^{oc}$ in the place of $\mathbf{Z}^{sc}$ in (16) and (17) results in solutions (pairs $k_{oc}$-$\omega$ and modes polarization) of the plate modes in the piezoelectric layer with OC and stress-free surfaces.

Another alternative form of the characteristic equations consists in using exclusively the electrical variables. Applying $\mathbf{t}^{\pm}{=}\mathbf{0}$ to eliminate $\mathbf{v}^{\pm}$ in (14) leads to an electrical system

$$\begin{bmatrix} \sigma \\ \sigma \end{bmatrix} \equiv \begin{bmatrix} \mathbf{E} \end{bmatrix} \begin{bmatrix} \psi^- \\ \psi^+ \end{bmatrix}, \text{ with } \mathbf{E} = \mathbf{Y} - \mathbf{X}(\mathbf{Z}^{sc})^{-1}\mathbf{K}. \tag{19a}$$

Elimination of $\psi^+$ from (19a) leads to a relation between $\sigma$ and $\psi^-$ via a scalar function, the so-called effective surface permittivity $\varepsilon_{eff}$ defined for the top surface by

$$\varepsilon_{eff}(\omega,k_1) \equiv \frac{\sigma}{-j|s_1|\psi^-} = \frac{E_{11}E_{22}-E_{12}E_{21}}{j|s_1|(E_{12}-E_{22})}. \tag{19b}$$

Here $E_{ij}$ are the elements of the $\mathbf{E}$-matrix. $\varepsilon_{eff}$ defined in (19b) is similar to the real part of the effective surface permittivity function of SAW problems[29] for all real $s_1{\equiv}k_1/\omega$ values except that it tends to infinity with $s_1 {\to} 0$ for all finite frequencies. Compared with the $\mathbf{Z}$-determinant, the $\varepsilon_{eff}$ function has the advantage of being sensitive only to the piezo-active modes. However, a pole and a zero of $\varepsilon_{eff}$, which are associated with the proper mode for respectively an SC and OC top surface, could be situated in extremely close proximity when the electromechanical coupling of the mode is poor.

## C. Multilayered plates

So far we have shown that guided waves in a piezoelectric layer can be obtained by first formulating the acoustic impedances and by second, reducing them to a SIM or an effective surface permittivity for one surface. The $\mathbf{Z}$-matrix for a single layer can be calculated directly from its definition (9) in terms of the associated spectral and modal matrices. But for a multilayer consisting of a finite number of different layers, writing out the overall $\mathbf{Z}$-matrix, which requires having recourse to a recursive procedure[9,16,17,19,21,22] is longer than writing out the overall transfer $\mathbf{P}$-matrix, which needs simple cascading (series multiplication) of the individual layers' $\mathbf{P}$-matrix. An alternative way consists in expressing the overall $\mathbf{Z}$-matrix in terms of the easy-processed overall $\mathbf{P}$-matrix using the following relations, which apply to a layered plate as well as to a single layer:

$$\mathbf{Z}_{11} = -\mathbf{P}_{21}^{-1}\mathbf{P}_{22}, \ \mathbf{Z}_{12} = \mathbf{P}_{21}^{-1}, \ \mathbf{Z}_{21} = \mathbf{P}_{12} - \mathbf{P}_{11}\mathbf{P}_{21}^{-1}\mathbf{P}_{22}, \text{ and } \mathbf{Z}_{22} = \mathbf{P}_{11}\mathbf{P}_{21}^{-1}, \quad (20a)$$

$$\mathbf{P}_{22} = -\mathbf{Z}_{12}^{-1}\mathbf{Z}_{11}, \ \mathbf{P}_{21} = \mathbf{Z}_{12}^{-1}, \ \mathbf{P}_{12} = \mathbf{Z}_{21} - \mathbf{Z}_{22}\mathbf{Z}_{12}^{-1}\mathbf{Z}_{11}, \text{ and } \mathbf{P}_{11} = \mathbf{Z}_{22}\mathbf{Z}_{12}^{-1}. \quad (20b)$$

Calculating the overall $\mathbf{Z}$-matrix from the overall $\mathbf{P}$-matrix after (20a) is viable and simpler than using the recursive algorithm as long as the numerical instability is absent. Otherwise, the recursive algorithm[17,22] has to be used to calculate the overall $\mathbf{Z}$-matrix. We shall not further deal with this aspect of the problem by assuming that the right SIM results we need are numerically available.

Now we return to the consideration of a unit cell in the Bragg coupler, which comprises of two layers of non-piezoelectric materials with high impedance-contrast. The mechanical contact at any interface is assumed to be perfect throughout this paper. We define the $\mathbf{Z}$ matrix for a unit cell in the same way as for a single layer, see (9), with at present $\mathbf{t}^{\pm}$ and $\mathbf{v}^{\pm}$ denoting the state vector evaluated at the top surface of the $SiO_2$-layer ($-$) and at the bottom surface of the W-layer ($+$). The same procedure applies to a periodic coupler consisting of $m$ unit cells, or to an arbitrarily multi-layered plate, when $\mathbf{t}^{\pm}$ and $\mathbf{v}^{\pm}$ refer to the state vector evaluated at the top ($-$) and bottom ($+$) surfaces of the whole structure. To stay general, we assume the coupler to be terminated with a known impedance $\mathbf{Z}_s$, which means that the relation $\mathbf{t}^{+} = \mathbf{Z}_s \mathbf{v}^{+}$ holds. Applying this relation to (9) leads to the expression of the SIM ($\mathbf{Z}_M$) defined for the coupler top surface :

$$\mathbf{t}^{-} \equiv \mathbf{Z}_M \mathbf{v}^{-}, \text{ with } \mathbf{Z}_M = \mathbf{Z}_{11} - \mathbf{Z}_{12}(\mathbf{Z}_{22} - \mathbf{Z}_s)^{-1}\mathbf{Z}_{21}. \quad (21)$$

In (21), $\mathbf{Z}_{ij}$ are the sub-matrices of the overall matrix $\mathbf{Z}$ referring to the whole coupler.

To determine the solutions proper to an AlN-Bragg plate with an electrically SC and stress-free surface, we apply the continuity condition at the AlN-coupler interface, expressed by $\mathbf{t}_c^{-} = \mathbf{t}_p^{+}$ and $\mathbf{v}_c^{-} = \mathbf{v}_p^{+}$. Here $\mathbf{t}_c^{-}$ and $\mathbf{t}_p^{+}$ (resp. $\mathbf{v}_c^{-}$ and $\mathbf{v}_p^{+}$) are the normal surface stress (resp. velocity) at the coupler and AlN side, respectively, of the interface. It yields

$$\mathbf{Z}_I \mathbf{v} = \mathbf{0}, \text{ with } \mathbf{Z}_I(\omega, k_1) = \mathbf{Z}_M - \mathbf{Z}^{+}. \quad (22)$$

In (22), $\mathbf{v}$ denotes the interface vibration velocity and $\mathbf{Z}^{+}$ is the SIM as determined in (16a) for an isolated AlN layer. When the coupler bottom surface is stress-free, $\mathbf{Z}_s=\mathbf{0}$ in the $\mathbf{Z}_M$-expression. Using $\mathbf{Z}_M = \mathbf{Z}_{11} - \mathbf{Z}_{12}\mathbf{Z}_{22}^{-1}\mathbf{Z}_{21}$ in the $\mathbf{Z}_I$-expression and then locating the $\mathbf{Z}_I$-determinant zeros, we can obtain the guided wave solutions (pairs $k_{sc}$-$\omega$) for a multilayer without substrate. A systematic computation yields the dispersion curves. The OC modes naturally result from replacing $\mathbf{Z}^{sc}$ with $\mathbf{Z}^{oc}$ in the matrix $\mathbf{Z}_\pm^\pm$, cf. (16-18).

**D. Effects of the resonator's electrodes and bottom substrate**

Now we show briefly how to include electrodes of finite thickness and a bottom substrate in the matrix formalism when dealing with a complete and more realistic SMR structure. Including the mechanical effects of finite thickness electrodes in the analysis complicates the formulation only to a small extent. For the inner electrode, a simple way is to incorporate its transfer matrix $\mathbf{P}_e$ into the coupler's one, and then to use the resultant matrix $\mathbf{P'}_m = \mathbf{P}_m\mathbf{P}_e$ instead of $\mathbf{P}_m$ everywhere. Another way of including the inner electrode, when making use of the recursive algorithm, consists in writing out first the sub-matrices of the inner electrode's $\mathbf{Z}$-matrix, say $\mathbf{Z'}_{ij}$, and then looking for the SIM $\mathbf{Z'}_M$ defined for its top surface with a known SIM $\mathbf{Z}_M$ exterior to its bottom surface. In this way, we easily obtain the expression of $\mathbf{Z'}_M$ as

$$\mathbf{t}_e^- \equiv \mathbf{Z'}_M \, \mathbf{v}_e^-, \text{ with } \mathbf{Z'}_M = \mathbf{Z'}_{11} - \mathbf{Z'}_{12} \, (\mathbf{Z'}_{22} - \mathbf{Z}_M)^{-1} \mathbf{Z'}_{21}. \qquad (23)$$

In (23), $\mathbf{Z}_M$ is the same as given in (21) referring to the coupler upper surface and including the substrate ($\mathbf{Z}_s$) when it is present, and $\mathbf{t}_e^-, \mathbf{v}_e^-$ refer to the state vector values calculated at the upper surface of the inner electrode. Comparison of (23) and (21) gives a hint at how the general SIM recursive relation looks. Namely, the contribution of an additional bottom part to the considered plate is expressed by subtracting the SIM ($\mathbf{Z}_s$, $\mathbf{Z}_M$) defined for the upper surface of the bottom part from the sub matrix ($\mathbf{Z}_{22}$, $\mathbf{Z'}_{22}$) to be inversed of the plate under consideration. As to the top electrode, let $\mathbf{Z}_e^+$ be the SIM defined for its bottom surface obtained by accounting for the zero stress at its upper surface. Instead of using $\mathbf{t}^- = \mathbf{0}$ in (15), we substitute the AlN-layer's upper surface stress $\mathbf{t}^-$ by $\mathbf{Z}_e^+ \mathbf{v}^-$. As a result, (16a) becomes

$$\mathbf{t}^+ \equiv \mathbf{Z}^+\mathbf{v}^+, \text{ with } \mathbf{Z}^+ = \mathbf{Z}_{22}^{sc} - \mathbf{Z}_{21}^{sc}(\mathbf{Z}_{11}^{sc} - \mathbf{Z}_e^+)^{-1} \mathbf{Z}_{12}^{sc}. \qquad (24)$$

Then, substituting $\mathbf{Z}^+$ in (22) with $\mathbf{Z}^+$ given by (24) allows the guided modes to be determined accounting for the top electrode.

As to substrate, $\mathbf{Z}_s = \mathbf{0}$ holds if the bottom substrate is the vacuum, and $\mathbf{Z}_s = \mathbf{Z}_{11}^p$, see (8), if it is a non-piezoelectric semi-infinite solid or a piezoelectric one with a SC surface. For the OC piezoelectric substrate, $\mathbf{Z}_s = \mathbf{Z}_{11}^{oc}$, see (18). In addition, the factor $g$ appearing in the expression of $\sigma^+$ in (10b) must be replaced with the element $Y_{11}$ calculated according to (8) for the substrate.

## IV. NUMERICAL RESULTS OF GUIDED MODES IN LAYERED PLATES

We numerically investigated structures consisting of a piezoelectric c-axis oriented AlN layer and a Bragg coupler having different numbers of layers W/SiO$_2$. Because of the coupler materials being isotropic and because of the special orientation of the AlN layer, the SH partial waves are decoupled from the SP ones and are piezo-inactive. In what follows, we focus only on the SP vibrations which are coupled with the electrical field. In our numerical examples, the thickness $h$ of AlN layer was taken as 1μm. For W (SiO$_2$), the thickness $H$ ($L$) was chosen as $H = 238.79$ nm ($L = 278.81$ nm), which corresponds to $\lambda/4$ when the wavelength $\lambda$ of the longitudinal thickness-mode in AlN is $\lambda = 2h$. First, we calculated the characteristic wave speeds for the three materials involved in our study. Table I lists the values we obtained with the physical constants we used in simulations. It shows that AlN is the fastest material and that W is the slowest one.

<div align="center">TABLE I</div>

Then, the acoustic spectra of Lamb waves in each individual layer were calculated. Though not reproduced here, they confirm that the Lamb and SH modes are uncoupled. We verified that the spectra of Lamb waves comprise both symmetric and anti-symmetric modes, and that as the frequency increases all of them tend to the vertically polarized shear bulk wave speed ($V_S$), except for the two lowest ones which tend to the SAW velocity ($V_{SAW}$) on the free surface of a half-space, which will be termed massive material in the subsequent text.

Second, we calculated the dispersion curves, as plotted in Fig. 2, for two different bi-layer plates, AlN/W and AlN/SiO$_2$. The former represents a high wave speed-contrast combination, and the latter is for the purpose of comparison with more complicated configurations. The electrical BC of the AlN layer was assumed to be SC in all numerical calculations. In addition to the form *f-k*, frequency as a function of wave number *k* (here named *k* instead of $k_1$), we also plotted the same curves in the form $V_p$-*f*, phase velocity versus frequency. All variables are in a normalized form ($f_n = f / f_0$, $f_0 = V_0/2h$), and $V_0 = 5882$ m/s is an arbitrarily chosen reference wave speed) for the purpose of more easily observing the asymptotic behavior of curves featured in plateau form. Two modes exist for all frequencies. The other modes (named higher order) do not appear below a corresponding threshold value $f_c$ (cut-off). None of modes is rigorously symmetric nor anti-symmetric because the bi-layer plates themselves do not exhibit any structural symmetry. Near the cut-off ($k \approx 0^+$), a few modes have a negative slope in the dispersion curve of *f-k* plot, a phenomenon similar to what happens to the so-called "anomalous Lamb modes".[30,31] The slope is usually considered as the group velocity ($V_g$) though some researchers consider that the original definition of $V_g$ is not applicable in this special *k*-range because the mode is amplitude-modulated in time.[31] As a general rule, the phase velocity $V_p$ of higher-order guided modes decreases with increasing frequency, with a limit corresponding to the speed $V_S$ of the vertically polarized shear bulk wave. With increasing frequency, the lowest branch goes up and then down after passing by a peak value of $V_p$, and finally tends to the SAW speed of the slower material (W or SiO$_2$). The next lowest branch goes directly down to the shear bulk speed of W for the AlN/W bi-layer. For the AlN/SiO$_2$ bi-layer, it first approaches the SAW speed of AlN ($V_p/V_0 \cong 0.9$) where it stays a while between $f_n = $ 2-3 near a horizontal line, termed plateau, before joining the shear bulk speed of massive SiO$_2$. Here we observe the mode repulsion phenomenon, i.e., the same branch of dispersion curves "shifts" above and below the plateau, as illustrated in Fig. 2 by the numbers 2 and 3. The wave patterns in this region resemble those of the non-dispersive SAW in massive AlN. The asymptotical lines for longitudinal BAW are not yet observable in the shown frequency range. For any mode exhibiting cut-off, the group velocity $V_g$ defined by the slope in *f-k* curve is zero at $f = f_c$. The interpretation is that a wave transversely resonant and sanding along $x_1$ ($k = 0$) transfers no energy in the $x_1$ direction.

<div align="center">FIG. 2.</div>

Figure 3 presents the acoustic modes spectra calculated for a 3-layer AlN/W/SiO$_2$ plate, a sandwich with the slowest layer embedded in between two relatively faster ones. As in bi-layer plates, two branches exist for all frequencies, and higher order branches appear for $f \geq f_c$, with a few of them exhibiting negative slope in the *f-k* plot of the dispersion curves near cut-off $f_c$. No mode in the full spectra is symmetric nor anti-symmetric. In the enlarged views of the dispersion curves $V_p$-$f$ plot, in a restricted velocity range but in more extended frequency range (up to $f_n = 25$), we can easily observe some plateaux where the wave speed seems to reach an asymptotic limit (horizontal line). An analysis of the origin of these peculiar behaviours enables one to get deeper physical insight of the wave motion in the layered structure. A plateau appears clearly in the $V_p$-$f$ plot as soon as $f_n \geq 2$, which is the speed of SAW in massive AlN. Two other plateaux exist, one above $V_p / V_0 = 0.5$ and one below $V_p / V_0 = 0.5$. The higher one is due to the SAW speed of massive SiO$_2$; however, the lower one is not due to the SAW speed in massive W. Contrary to what is expected, the lowest branch in this structure approaches a wave speed that is not the SAW velocity, neither in any of the two external materials SiO$_2$ and AlN, nor in the middle W. A careful analysis reveals that it tends to the wave speed of the interfacial (Stoneley) mode which would exist and propagate near the interface of W and SiO$_2$ when both fill up half-spaces. In addition, we observe another asymptotic limit at $V_p / V_0 \cong 0.5$ for $f_n > 10$, which is the shear bulk wave in massive W. At higher frequencies ($f_n > 15$), some branches reach a plateau-like zone just below $V_p / V_0 = 0.9$ over a finite frequency range, which is due to the longitudinal bulk wave of massive W; the branches then undergo a sharp decrease in phase velocity, after which they approach first the asymptotic limit of shear bulk speed of massive SiO$_2$ ($V_p / V_0 \approx 0.63$) and then the shear bulk speed of massive W ($V_p / V_0 \approx 0.5$). These results clearly show the acoustic confinement in the slow layer (W) of the guided modes at very short wavelength regime. In addition to one interface and two surface modes, two families of guided modes can be distinguished, one for those tending to the shear bulk speed of W ($V_p / V_0 \approx 0.5$) and the other for those approaching the shear bulk speed of SiO$_2$ ($V_p / V_0 \approx 0.63$).

FIG. 3.

To go further, we have also calculated the dispersion curves for a 4-layer AlN/W/SiO$_2$/W plate, as shown in Fig. 4. The characteristics of higher modes cut-off, no symmetry, and negative slope are similar to the previously presented 2- and 3-layer plates, except that here the number of modes is larger (58 against 47 in Fig. 3, and 36 for both AlN/SiO$_2$ and AlN/W configurations in Fig. 2) within the same frequency range ($f_n \leq 15$). The lowest branch tends to the SAW speed of massive W because W is now one of the outside layers. The next lowest branch approaches rapidly (at $f_n \geq 2$) the speed of the interfacial mode. Another branch also tends to the interface mode speed, but at a much higher frequency, $f_n \geq 8$. This phenomenon is logically explained by the presence of two W/SiO$_2$ interfaces in the current configuration-piezoelectric AlN combined with a 3-layer coupler W/SiO$_2$/W. The branch which reaches the interfacial wave speed first is essentially at the interface of SiO$_2$ with the outside W-layer, while the other one is mainly at the interface of SiO$_2$ with the embedded W-layer. No mode tends to the SAW speed of massive SiO$_2$ because this layer has no stress-free surface. The plateau above $V_p / V_0 = 0.9$ is attributed to the SAW in massive AlN material as in Fig.

3. To observe the guided modes near the bulk wave speeds in massive materials, it is necessary to go further in frequencies.

FIG. 4.

Figure 5 presents the dispersion curves for $f_n$ up to 50 along with all asymptotic lines associated with a characteristic wave speed. In sufficiently short wavelength regime, the guided modes can be classified into 2 families according to their asymptotic behavior. The family with lower speeds reaches by pair the shear bulk speed ($V_p / V_0 \approx 0.5$) of massive W, a phenomenon due to the presence of two separate layers of W in the structure. The family with higher speeds goes singly to the shear bulk speed ($V_p / V_0 \approx 0.63$) of massive $SiO_2$, as expected when only one layer is made of $SiO_2$ material in the whole structure. We also observe a double-mode plateau around $V_p / V_0 \approx 0.9$, the speed of longitudinal bulk wave in massive W, which can be interpreted as an intermediate wave confinement in either W-layer. The guided modes near these plateaux have almost the same properties as the classical shear or longitudinal bulk waves in unbounded W material. The longitudinal BAW speed in $SiO_2$ is a little higher than the line 7, and the longitudinal BAW in AlN is too high to be shown in this graph.

Numerical instabilities were appearing for $f_n$ higher than 10 around $V_p / V_0 = 1$ during computations using the transfer matrix. No instability was observed when using the impedance matrix formalism al though some modes could not be determined (missing data) using the SIM defined for one specific surface. However, thanks to the SIM approach which is unconditionally stable irrespective of the total number of layers and individual layer thickness, in contrast to the well known TMM, it suffices to repeatedly apply the SIM at different locations within the layered structure for the complete wave spectra of all guided modes to be determined, even in extremely short wavelength regime. This mainly concerns regions where guided modes tend to SAW or bulk wave speeds of a component material. Relatively straightforward and efficient from the computational point of view, the approach is also flexible because the characteristic equation, always in a form of $|Z_l^m - Z_u^{m+1}| = 0$, can be written for an arbitrary interface as well as for an external surface. This flexibility allows the numerical results to be obtained for the dispersion curves and field distributions by selecting the only interface in the neighbourhood of which mode confinement occurs or the electromechanical energy flux is concentred when the full set of solutions is not required.

FIG. 5.

## V. CONCLUSIONS AND FURTHER DISCUSSION

Calculations of guided waves in layered plates mixing piezoelectric, dielectric and metallic layers have been developed using the Stroh formalism and matrix presentation. The characteristic equations for the dispersion curves are derived in two forms: an acoustic surface impedance matrix (SIM) and an electrical scalar function. The SIM is expressed in a unified elastic form for both piezo and non-piezo materials, which makes easy the application of the field continuity at the interfaces of heterostructures. By repeatedly applying the unconditionally stable SIM formalism at different interface loca-tions, we have been able to obtain the full set of solutions, even for frequencies up to several tenths of the fundamental thickness mode resonant frequency. Numerical investigations for plates having up to four layers and three materials show an extreme complexity of the acoustic spectra. They include 1) the slope of *f-k* curves near the cut-off, for a given mode, can be negative or positive depending on the specific stack configuration, and more than one mode exhibit this feature. This suggests that the sign of the dispersion curve slope of certain modes can be controlled by layers' stack order and/or relative thickness ratio; 2) interfacial waves exist at short wavelength regime between the W and $SiO_2$ layers, but they do not appear at the interface between AlN and W nor between AlN and $SiO_2$ layers. As a general rule, this depends on the velocity ratio; 3) the energy of all guided acoustic waves is confined in the slow W-layer when the wave length is short enough, though intermediate confinement occurs in the $SiO_2$ layer when it is embedded. The asymptotic behavior of the dispersion curves is physically originated from the proper modes of BAW and SAW in massive materials of the constituent materials. The mode clustering and forbidden regions are not yet observed because the number of layers is still low and the structural periodicity is insignificant. With an increase in the number of unit cells, say for $N \geq 3$, one can expect the appearance of pseudo band gaps in the wave spectra due to Bragg effects.

Investigation of Lamb modes in SMR including two metal electrodes and a Si substrate are under way. The effects of the semi-infinite substrate and thick electrodes, already included in the general expressions derived in Sec. III-D, are to be accounted for in future numerical simulations. An extension of the analysis to the purely imaginary domain as well as to the complex domain of the wave number is needed in order to have a complete knowledge of the full Lamb wave spectra, which becomes indispensable for modeling the lateral propagation phenomena[33,34] and for understanding the spurious signals observed in SMR-based filter responses. Knowledge of the energy trapping properties and wave motion patterns, if desired, can be gained by examining the through-thickness distribution of electro-acoustic fields for any specific mode defined by the pair value of $\omega$-$k_1$. The most direct effects of the electrodes (of material Mo) are a considerable lowering of the proper resonant frequencies of the piezoelectric AlN resonator. The presence of a bottom substrate ($\mathbf{Z}_s \neq \mathbf{0}$) leads to solutions of guided surface modes which are multiple and dispersive, similar to the conventional Rayleigh SAW, provided that the substrate is a faster material and the wavelength is comparable to the overall thickness of AlN layer added to the coupler. In the considered configuration (Fig. 1), this type of solutions only exist for very low frequencies. In normal resonator operation, most of the solutions are of leaky SAW type which pertain more or less bulk radiation into the substrate depending on the coupler parameters and frequency. As a consequence, the wave-number of any mode having a phase speed faster than the SV-polarized BAW in the substrate is necessarily complex, leading to an attenuation as they propagate along the surface (in $x_1$). At short wavelength regime ($\lambda < h/5$) and for a given $\omega$-$k_1$ pair, the wave pattern in each layer tends to be independent one from the other. As a result, one cannot talk about the mode type (SAW- or Lamb-like) for the overall structure, which has no longer a precise signification. The exact wave pattern in a heterostructure is easy of access in all cases by examining the field profiles as a function of the thickness position (for a fixed $\omega$-$k_1$ pair).

**FIGURE CAPTIONS**

FIG. 1. A layered plate composed of a piezoelectric AlN layer with *N* identical cells of bi-layer W/SiO$_2$. Also shown are the coordinates system and the relevant physical quantities at the surface of a layer. Electric variables are not involved for non-piezoelectric layers (Color online).

FIG. 2. Dispersion curves of guided waves in two bi-layer plates, AlN/W (left) and AlN/SiO$_2$ (right). The top panes are the overall views of frequency against wave number, both normalized, and the bottom panes are enlarged views of the phase velocity versus frequency to show the cluster and plateau behaviour of guided modes in some dispersion regions. Blue lines: SAW in massive AlN; magenta lines: SAW in the other material; red lines: shear BAW in the slower material; green lines: shear BAW in AlN (Color online).

FIG. 3. Dispersion curves of guided waves in a 3-layer AlN/W/SiO$_2$ plate: on the left is the normalized  frequency $f_n$ versus the wave-number $k$ in a limited range, on the right are two enlarged views of the phase velocity $V_p$ as a function of the frequency $f_n$, within a restricted $V_p$-range and an extended $f_n$-range. The curve branch with the slowest speed ($V_p / V_0 \cong 0.479$) is the interface mode existing near the interface of W and SiO$_2$ layers and having essentially the same properties as the one which would exist and propagate at the interface between these two materials when they are massive half-spaces (Stoneley waves).

FIG. 4. Dispersion curves of guided waves in a 4-layer AlN/W/SiO$_2$/W plate: on the left is the normalized  frequency $f_n$ versus the wave-number $k$ in a limited range, on the right are two enlarged views of phase velocity $V_p$ plotted as a function of frequency $f_n$ within a restricted $V_p$-range and an extended $f_n$-range. The curve branch with the slowest speed ($V_p / V_0 \cong 0.454$) is the SAW existing near the exterior W-layer surface. Two branches tend to the same wave speed of the interface mode because there are two interfaces of W-SiO$_2$.

FIG. 5 Dispersion curves of the same plate as in Fig. 4, showing the phase velocity V$_p$ versus the frequency *f* for a more extended *f*-range, to show their asymptotic limits. The line numbered 1 is for SAW in massive W; the line 2 is for interface waves (two branches merging into one); the line 3 is for shear BAW in massive W (a group of two modes, one in the inner W and the other in the outer W layer); the line 4 is for shear BAW in massive SiO$_2$; the line 5 is longitudinal wave in massive W; the lines 6 and 7 are for SAW and BAW in massive AlN, respectively.

**TABLE CAPTION**

TABLE I – Characteristic wave speeds of the three materials used in the multilayers.

TABLE I

|  | $V_L$ (m/s) | Y-polar-$V_S$(m/s) | $2^{nd}$ $V_S$ (m/s) | $V_{SAW}$ (m/s) |
|---|---|---|---|---|
| c-axis AlN: SC/OC | 10605/10939 | 5796 | 5796 | 5394/5402 |
| X-propagation | 9911 (non-piezo) | 5597 | 5796/5867 | 5417/5437 |
| W (High-impedance) | 5224 | 2887 |  | 2669 |
| SiO$_2$ (Low-impedance) | 6099 | 3655 |  | 3342 |

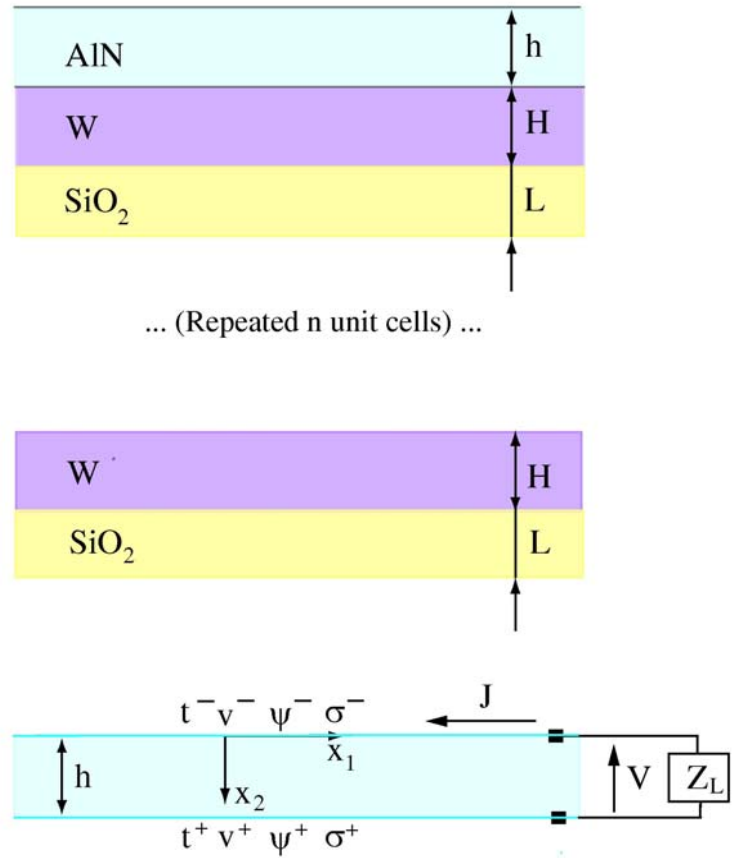

FIG. 1.

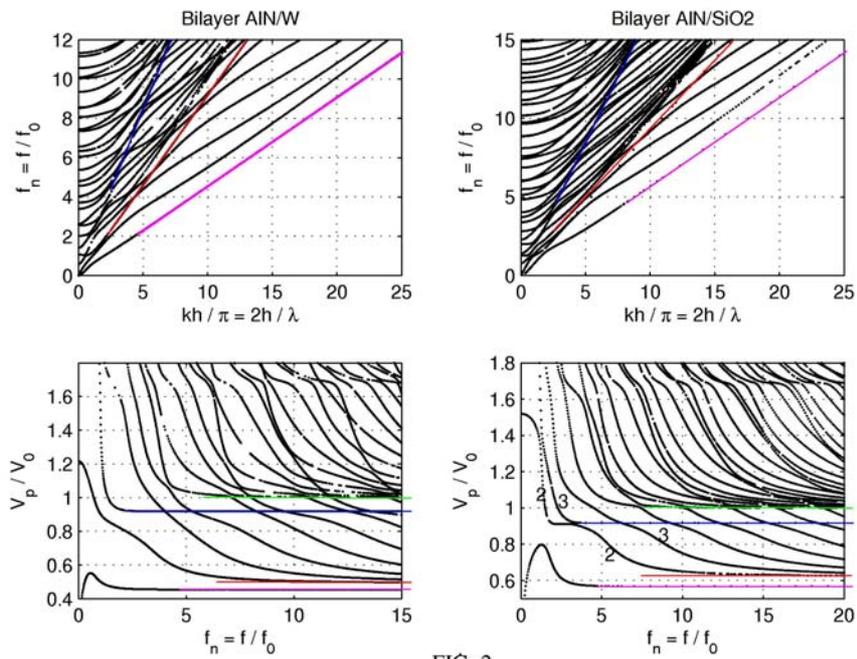

FIG. 2.

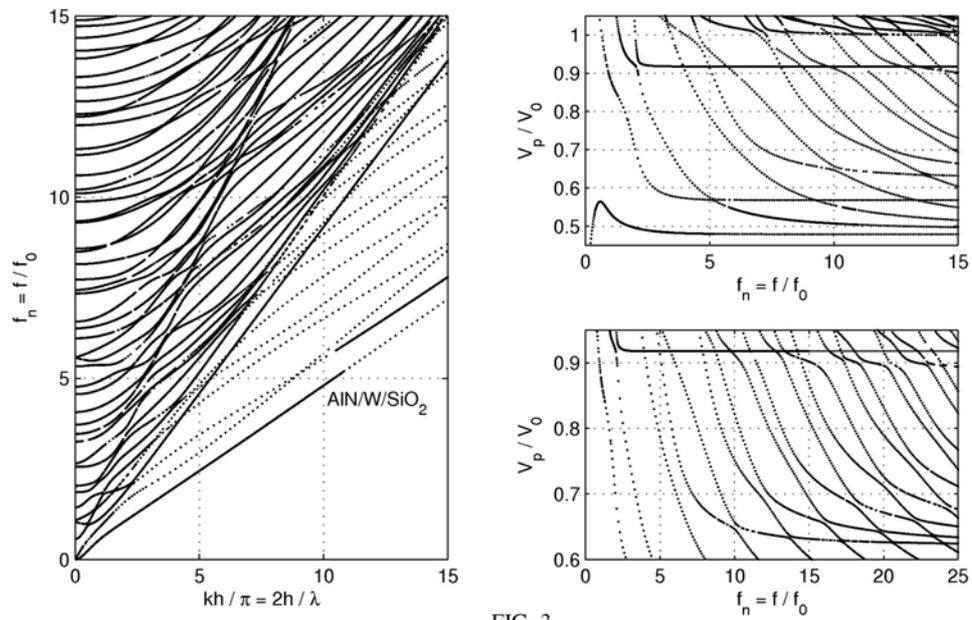

FIG. 3.

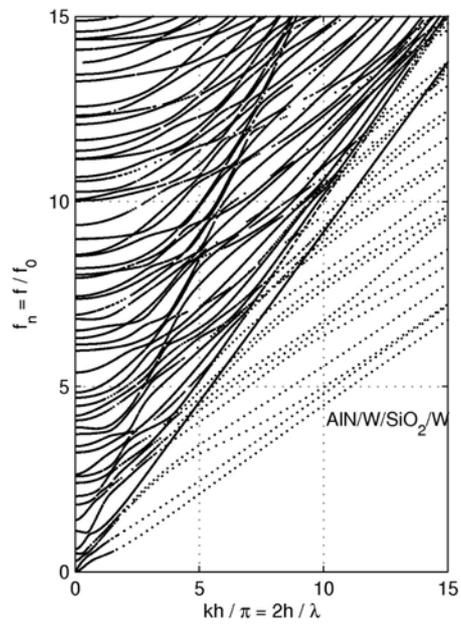

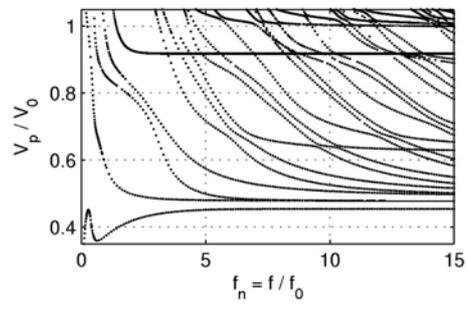

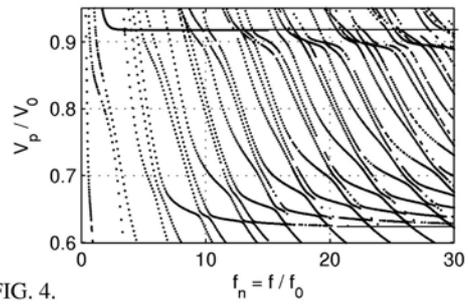

FIG. 4.

FIG. 5.

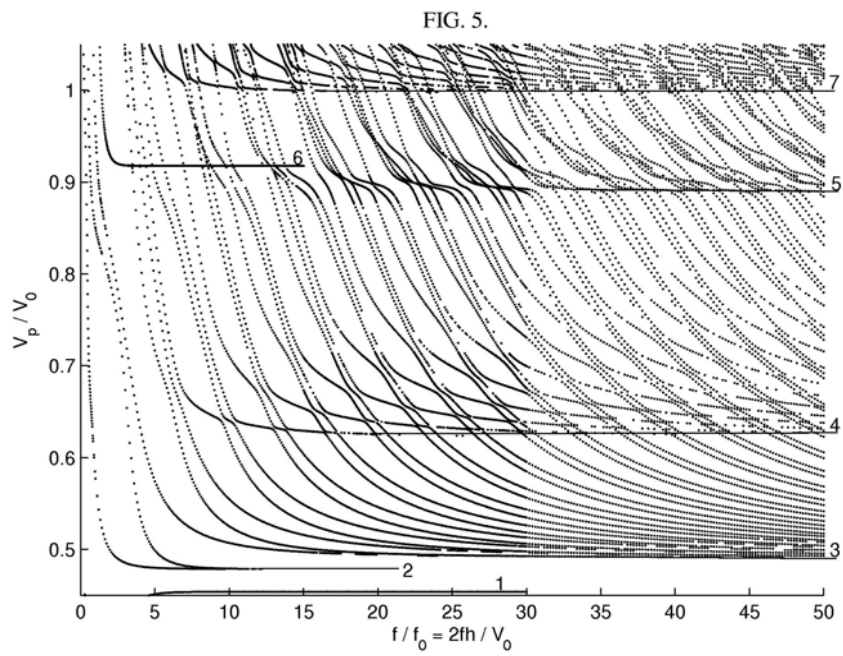